\documentclass[twocolumn,tighten]{aastex62}

\newcommand{\kms}{km\ s$^{-1}$}

\graphicspath{{./}{figures/}}

\received{}
\revised{}
\accepted{\today}
\submitjournal{ApJL}

\shorttitle{DDO68 C: the actual appearance of a ghost satellite dwarf}
\shortauthors{Annibali et al.}

\begin{document}

\title{DDO68 C: the actual appearance of a ghost satellite dwarf through adaptive optics at the Large Binocular Telescope}

\correspondingauthor{Francesca Annibali}
\email{francesca.annibali@inaf.it}

\author{Francesca Annibali}
\affil{INAF - Osservatorio di Astrofisica e Scienza dello Spazio \\ 
Via Piero Gobetti, 93/3 \\ 
I - 40129 Bologna, Italy}

\author{Enrico Pinna}
\affil{INAF - Osservatorio Astrofisico di Arcetri \\ 
Largo Enrico Fermi 5 \\ 
I - 50125 Firenze, Italy}

\author{Leslie~K. Hunt}
\affil{INAF - Osservatorio Astrofisico di Arcetri \\ 
Largo Enrico Fermi 5 \\ 
I - 50125 Firenze, Italy}

\author{Diego Paris}
\affil{INAF - Osservatorio Astronomico di Roma \\ 
via Frascati 33 \\ 
I - 00078 Monte Porzio Catone, Italy}

\author{Felice Cusano}
\affil{INAF - Osservatorio di Astrofisica e Scienza dello Spazio \\ 
Via Piero Gobetti, 93/3 \\ 
I - 40129 Bologna, Italy}

\author{Michele Bellazzini}
\affil{INAF - Osservatorio di Astrofisica e Scienza dello Spazio \\ 
Via Piero Gobetti, 93/3 \\ 
I - 40129 Bologna, Italy}

\author{John M. Cannon}
\affil{Macalester College \\ 
1600 Grand Ave., Saint Paul \\ 
MN, 55105, USA}

\author{Raffaele Pascale}
\affil{INAF - Osservatorio di Astrofisica e Scienza dello Spazio \\ 
Via Piero Gobetti, 93/3 \\ 
I - 40129 Bologna, Italy}

\author{Monica Tosi}
\affil{INAF - Osservatorio di Astrofisica e Scienza dello Spazio \\ 
Via Piero Gobetti, 93/3 \\ 
I - 40129 Bologna, Italy}

\author{Fabio Rossi}
\affil{INAF - Osservatorio Astrofisico di Arcetri \\ 
Largo Enrico Fermi 5 \\ 
I - 50125 Firenze, Italy}



\begin{abstract}

Through adaptive optics (AO) imaging with the SOUL+LUCI instrument at the Large Binocular Telescope we were able to resolve, for the first time, 
individual stars in the gas-rich galaxy DDO68~C. This system was already suggested to be interacting with the extremely metal poor dwarf DDO68, but its nature has remained elusive so far because of the presence of a bright foreground star close to its line of sight, that hampers a detailed study of its  stellar population and distance. In our study, we turned this interloper star into an opportunity to have a deeper insight on DDO68~C, using it as a guide star for the AO correction. 
Although the new data do not allow for a direct distance measurement through the red giant branch tip method, 
the combined analysis of the resolved-star color-magnitude diagram, of archival GALEX FUV and NUV photometry, and of H$\alpha$ data 
provides a self-consistent picture in which  DDO68~C is at the same $\sim$13 Mpc distance as its candidate companion DDO68. 
These results indicate that DDO68 is a unique case of a low mass dwarf, less massive than the Magellanic Clouds,  interacting with three satellites (DDO68~C and two previously confirmed accreting systems), providing useful constraints on cosmological models and 
a potential explanation for its  anomalous extremely low metallicity.

\end{abstract}



\section{Introduction} \label{intro}

The most metal-poor, star-forming dwarf galaxies in the nearby Universe are of paramount importance since they 
offer the opportunity to study the details of star formation and chemical evolution in a regime similar to that of 
primordial galaxies in the early Universe \citep[e.g.,][]{Izotov94,KO2000,Senchyna2019,McQuinn2020, AnnibaliTosi22}. 
Extremely metal poor dwarfs (XMPs), with an oxygen abundance of 12+log(O/H)$\leq$7.35 or $\lesssim$4 percent solar \cite[e.g.,][]{Guseva15}, are quite rare galaxies, 
and because of their interest they have been actively sought in recent years by many groups \citep{Pustilnik05,Izotov12,Skillman2013,Hirschauer2016,Guseva2017,Hsyu2017,Yang2017,Hsyu2018,Senchyna2019,Kojima2020,McQuinn2020,Pustilnik2021}.
 While some XMPs can be explained with the combined effects of inefficient star formation and high metal loss via galactic winds at low galaxy masses \citep[e.g.,][]{Skillman2013,Hirschauer2016}, other relatively massive XMPs are quite difficult to understand and challenge models of galaxy evolution \citep{Izotov2018,McQuinn2020}. These systems are strong outliers in the luminosity-metallicity and mass-metallicity relations
 \citep[e.g.][]{Mannucci2011,Berg12,Hunt2016}, with a measured metallicity far too low for their stellar mass/luminosity. Accretion of metal poor gas either from the intergalactic medium \citep{Filho2015} or from interaction with smaller companions \citep{Ekta2010,McQuinn2020} are among the mechanisms proposed to explain their anomalous low metal content, but are not always sufficient \citep[e.g.,][]{Matteucci85,Marconi94,Pascale2022}.

An iconic example of an XMP with discrepant metallicity is the nearby \citep[D$\sim$13 Mpc,][]{Tikhonov2014,Sacchi16} dwarf irregular galaxy DDO68,  whose HII regions' oxygen abundance of just $\sim$3\% solar \citep{Pustilnik05,Annibali2019b} appears incompatible with its stellar mass of  M${_\star}\sim10^8$ M$_{\odot}$ \citep{Sacchi16}. Although located in a Void \citep{Pustilnik11},  DDO68 has been suggested to have merged with a 10-20 times smaller gas-rich companion  (named DDO68~B), responsible for its very distorted morphology characterized by a large ``cometary tail'' \citep[][]{Tikhonov2014,Sacchi16}. Deep photometry with HST and the Large Binocular Telescope (LBT) revealed the presence of a second, even smaller (M${_\star}\sim10^6$ M$_{\odot}$) interacting satellite  (named S1), likely a gas-free spheroidal galaxy \citep{Annibali2016,Annibali2019a}. Consequently, DDO68 is so far the smallest dwarf galaxy with clear evidence for accretion of at least two satellites.

\cite{Pascale2022} developed detailed hydro-dynamical N-body simulations of the merger of DDO68 with its two smaller satellites to reproduce the observed stellar and gaseous morphologies and the HI kinematics. They concluded that gas dilution from a twenty times smaller, metal-poor, gas-rich satellite, such as DDO68~B, is unable to explain the extremely low chemical abundances derived in DDO68; however, metallicities almost as low as the observed ones are reproduced by the simulations assuming that the companion's accreted material has been inefficiently mixed with that of the host galaxy and that the observed HII regions trace in fact the composition of the acquired satellite rather than that of DDO68’s main body.

\begin{figure}
\includegraphics[width=\columnwidth]{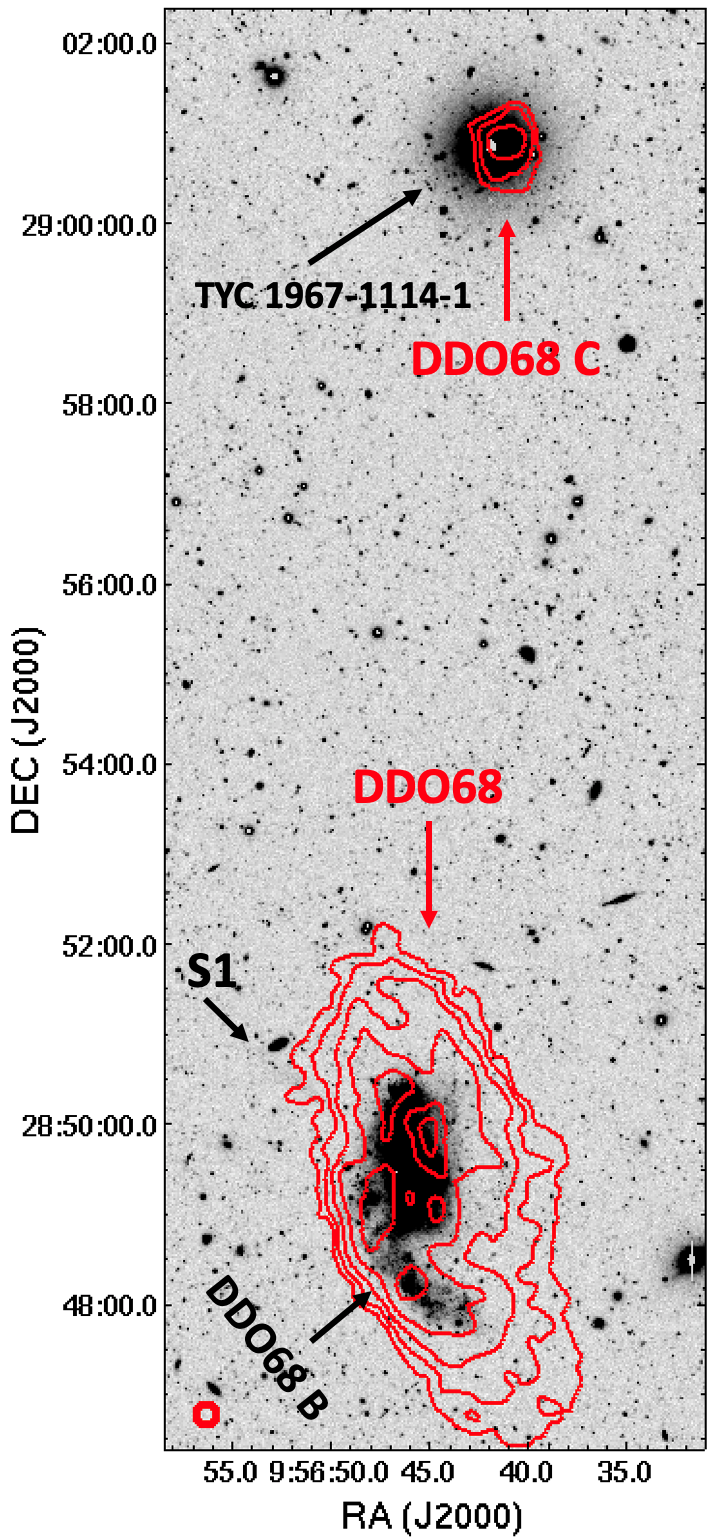}
    \caption{LBT g image of DDO68 from \cite{Annibali2016} with superimposed HI contours 
    at (1.25, 2.5, 5, 10, 20) $\times$ 10$^{20}$\ cm$^{-2}$ \citep[VLA data from][]{Cannon2014}. 
    DDO68~C is detected in HI $\sim$11$^{\prime}$ north from DDO68, but is not visible in the optical because it is hidden by the bright foreground star TYC~1967-1114-1. The low surface brightness HI bridge connecting DDO68 and DDO68~C is detected in lower resolution HI data but not visible here.  The other two confirmed satellites DDO68~B and S1 are also indicated. The 
    15$^{\prime\prime}$ beam of the HI data is shown at the bottom left.}
    \label{fig_intro}
\end{figure}

\begin{figure*}
\begin{center}
\includegraphics[width=\textwidth]{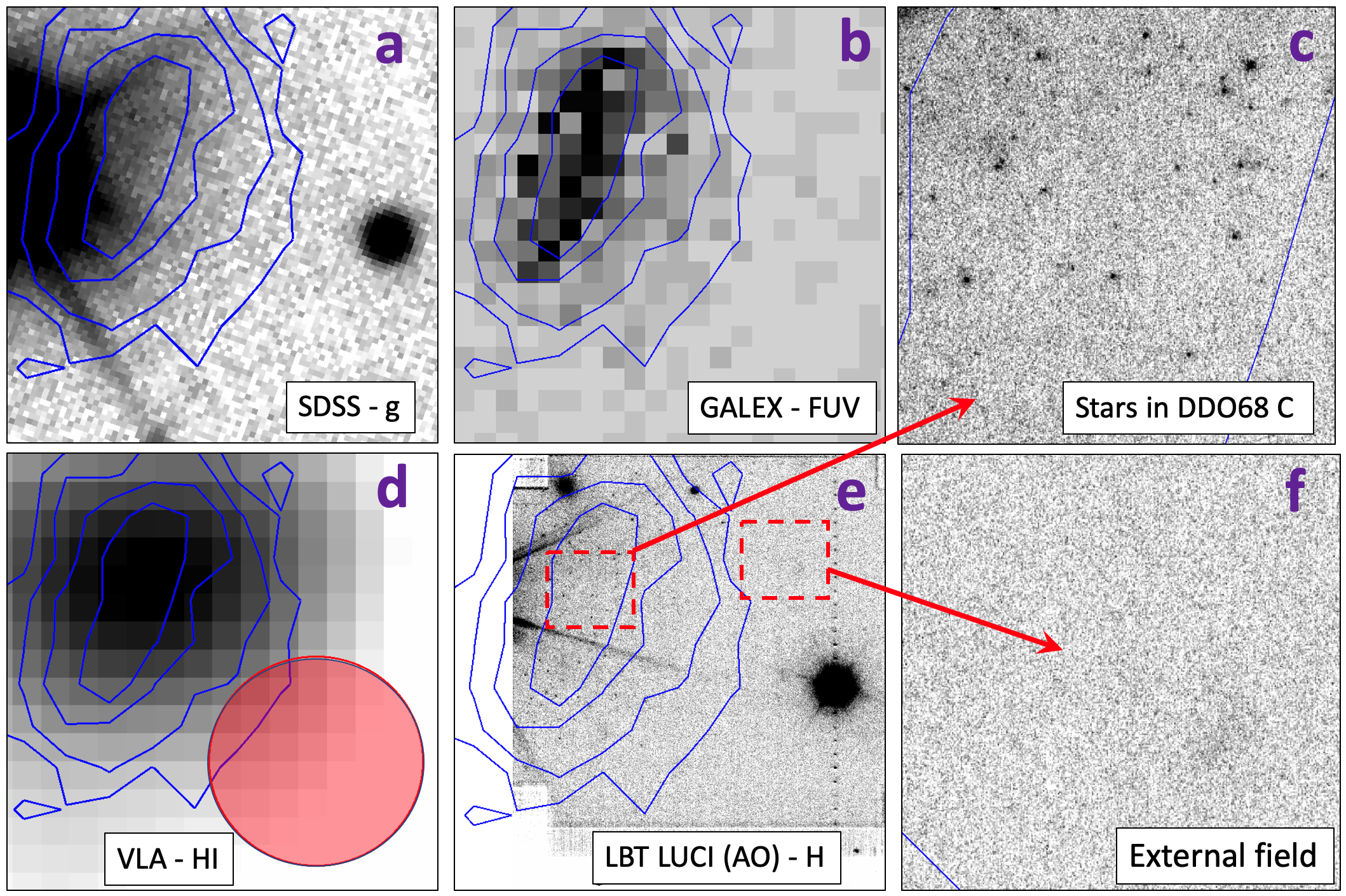}
   \caption{Images of DDO68~C in different bands: in panels a, b, d and e we show 30$^{\prime\prime}\times30^{\prime\prime}$ portions of the SDSS g image, the GALEX FUV image, the VLA HI image (with the 15$^{\prime\prime}$ beam), and the SOUL+LUCI AO H-band image acquired with the LBT, respectively. FUV contours at $\sim$0.3, 0.6, 1.2, and 2.3 $\times$10$^{-15}$ erg \ s$^{-1}$ \ cm$^{-2}$ are overlaid in blue on all images. 
   In the SDSS image, the galaxy is hidden by the bright  foreground red star TYC~1967-1114-1, but is visible in FUV and in HI.  The bright source 
   to the right of DDO68~C in panels a and e is the foreground star 2MASS~J09563976+2900457. Panels c and f are 
   6$^{\prime\prime}\times6^{\prime\prime}$  portions of the LUCI image in the H band centered on the stellar 
   emission from DDO68~C and on an outer field where no stars associated to DDO68~C are detected.}
   \label{fig_multi}
\end{center}
\end{figure*}

An alternative solution to the problem of the low metallicity in DDO68 was proposed by \cite{Cannon2014} based on the 
discovery of a candidate third interacting system in Karl Jansky Very Large Array (VLA) 21-cm observations. 
This gas-rich (M$_{HI}\sim2.8\times10^7$M$_{\odot}$) system, named DDO68~C (see Fig.~\ref{fig_intro} and Table~\ref{tab1}), lies at an angular separation of just 11$^{\prime}$ from DDO68, has its same systemic velocity, and appears connected to DDO68 by a low surface brightness HI bridge. These properties could indicate that the two galaxies are located at about the same distance, separated by just $\sim$40 kpc in projection, and that they are gravitationally interacting.
Unfortunately, both the stellar population properties and the distance of DDO68~C remain poorly constrained due to the presence of the  bright foreground  star  TYC~1967-1114-1 (with spectral type of K0V-K2V and magnitude of K=8.5), close to the galaxy line of sight (Fig.~\ref{fig_intro} and Fig.~\ref{fig_multi}). This bright red interloper hampers the detection of DDO68~C from the ground in seeing-limited mode. On the other hand, GALEX FUV and NUV images, where the red foreground star is relatively faint, reveal a diffuse emission co-located with the HI detection. This suggests the presence of relatively young stars, but can not provide direct information on the galaxy's distance.
While the low surface brightness HI gas that connects DDO68 and DDO68~C offers tantalizing evidence for an ongoing interaction, a direct measurement of DDO68~C's distance would be important to place the physical association of the two systems on a more secure footing.

\begin{figure*}
\begin{center}
\includegraphics[width=\textwidth]{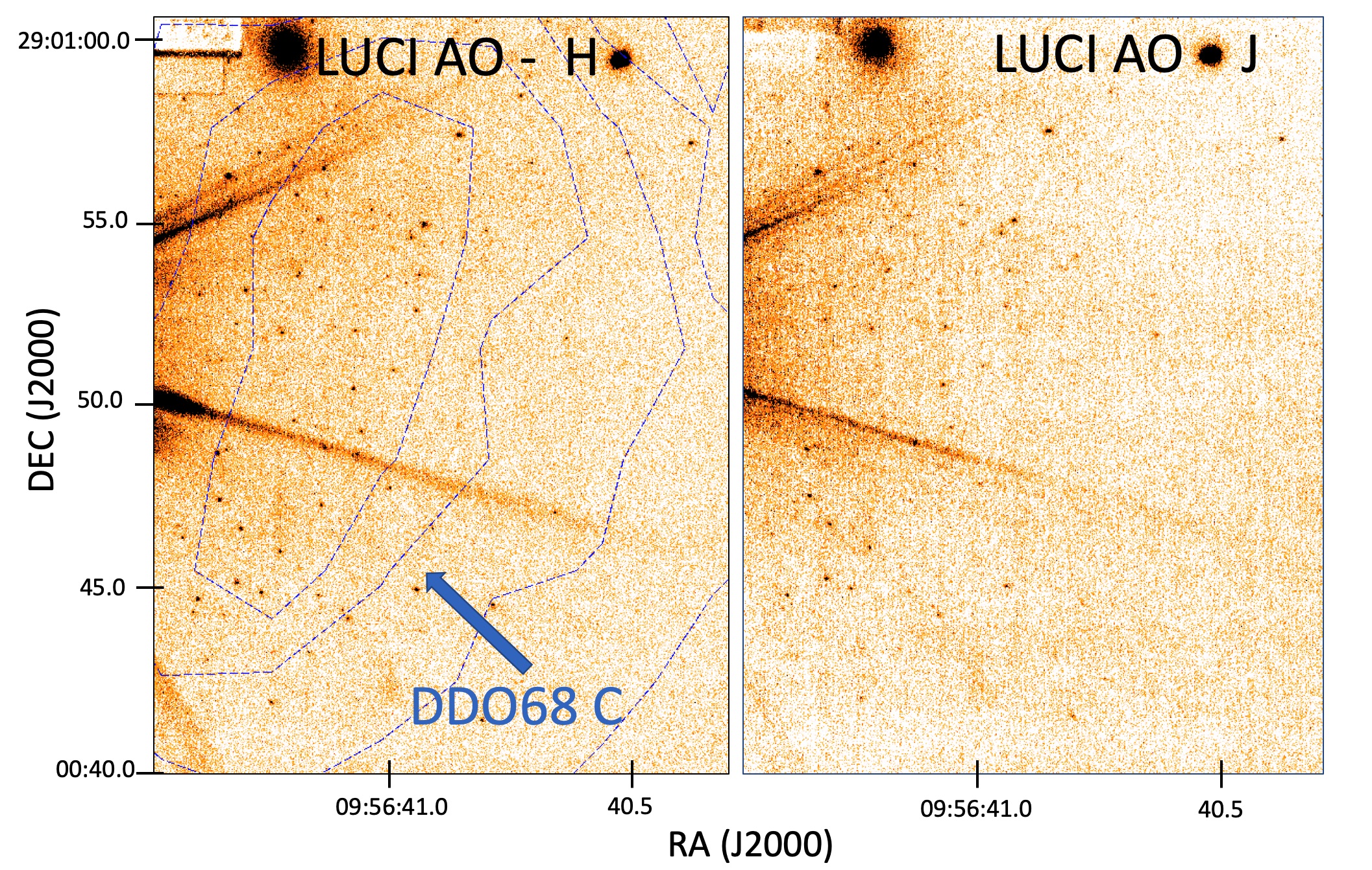}
    \caption{15$^{\prime\prime}\times$30$^{\prime\prime}$  portions of the larger SOUL+LUCI AO images of DDO68~C in the H (left) and J (right) bands capable of resolving its individual stars. Superimposed to the H image are the same GALEX FUV contours as in Fig.~\ref{fig_multi}.
    The bright foreground star TYC~1967-1114-1 used for the AO correction 
    is outside the LUCI field of view to the East, but its prominent spikes are visible in the images. The fainter 2MASS~J09563976+2900457 star used to evaluate the point spread function and to calibrate the photometry is to the West, outside the 
    displayed image portion.}
    \label{fig_lbt}
    \end{center}
\end{figure*}

To make progress in understanding the nature of DDO68~C, we exploited the high spatial resolution offered by adaptive optics (AO) and  acquired new deep imaging of the galaxy in the J and H bands with the SOUL+LUCI instrument on the LBT using the bright interloper star TYC~1967-1114-1 for the AO correction. This paper presents our new data capable of resolving, for the first time ever, individual stars in DDO68~C. The observations and the data reduction procedure are described in Section~2, while we present the resolved-star color-magnitude diagram (CMD)  
in  Section~3. The discussion of the results and our conclusions are given in Section~4.

 \begin{table}
        \centering
        \caption{Properties of DDO68~C. The last column reports the reference papers: 1=\citet{Cannon2014}; 2=\citet{Pascale2022}; 3=this work. 
        ($\star$)=for an assumed distance of $\sim$13 Mpc.
        \label{tab1}}
        \label{tab:props}
        \begin{tabular}{l c c}
        \hline\hline
        Property                       & Value                                         & Ref.\\
        \hline
        RA                             &        {\rm {09$^h$ 56$^m$ 41.$^s$07}}                         & 1   \\
        DEC                            &       +29$^{\circ}$ 00$^{\prime}$ 50.$^{\prime\prime}$74 & 1   \\
        Systemic velocity              & $\sim$ 506 \kms                                    & 1 \\
        Rotation velocity              & $\gtrsim$ 7.5-10 \kms                              & 1 \\
        Dynamical mass$^{\star}$                 & $\gtrsim$  10$^8$M$_{\odot}$                  & 2   \\
         HI mass$^{\star}$                           & (2.8$\pm$0.5)$\times10^7$M$_{\odot}$               & 1   \\
        SFR from FUV$^{\star}$               & (1.4$\pm$0.4)$\times$10$^{-3}$ M$_{\odot}$  yr$^{-1}$          & 1 \\
        H$\alpha$ luminosity$^{\star}$             & $<$8.5$\times$10$^{36}$erg \ s$^{-1}$         & 1 \\
        SFR from H$\alpha$$^{\star}$               & $<$7$\times$10$^{-5}$ M$_{\odot}$ yr$^{-1}$          & 1 \\
        m$_{FUV}$                      & 19.55 $\pm$ 0.01  [AB mag]                     & 3 \\
         m$_{FUV}$ - m$_{NUV}$         & 0.21 $\pm$ 0.06  [AB mag]                     & 3\\
        \hline
        \end{tabular}
\end{table}

\section{Observations and Data Reduction} \label{observations}

Observations were performed at the LBT with the near infrared LUCI camera in single conjugate adaptive optics (SCAO) mode 
with the second generation AO instrument SOUL \citep{Pinna2016,Pinna2021}. 
Images of DDO68~C were obtained in the J and H bands during two separate runs on January 2021 and March 2021 performing
exposures of 120 s duration (8 s exposure $\times$ 15 frames) dithered according to a rectangular pattern with $\sim$4$^{\prime\prime}$ maximum displacement in the X or Y directions.  Additional exposures were acquired at positions $\sim$40$^{\prime\prime}$ off the science target for background evaluation. The AO system was correcting 500 modes at a framerate of 1200 Hz. 

The total exposure times on target, thus excluding sky observations, are 8280 s in H and 7920 s in J. The bright K=8.5 mag foreground star 
TYC~1967-1114-1, displaced by $\sim$10$^{\prime\prime}$ from DDO68~C's FUV centroid, was used for the AO correction. Observations were performed under average seeing conditions of 0.7$^{\prime\prime}$ - 1.0$^{\prime\prime}$. FWHMs in the AO images are in the range $\sim$0.08$^{\prime\prime}$  to $\sim$0.25$^{\prime\prime}$ in J and $\sim$0.06$^{\prime\prime}$  to $\sim$0.2$^{\prime\prime}$  in H.  

The individual exposures were calibrated and sky-subtracted using a dedicated pipeline developed at INAF - Rome Observatory to reduce LBT LUCI imaging data.  The preliminary reduction procedure consists of removal of the dark current and of flat field correction. 
The dark removal is accomplished by subtracting from each (both scientific and sky) image a median stacked dark image (masterdark) obtained by combining a set of dark frames acquired with the same detector integration time (DIT) and DIT number (NDIT) as the scientific dataset. 
Each dark-subtracted image is then divided by a median stacked flat image (masterflat) obtained through the combination of a set of dark-subtracted flat images.
The pre-reduced science images are then sky-subtracted using a mean sky image obtained by averaging the two pre-reduced sky images 
acquired at the bounds of the temporal sequence of the scientific dataset.

Since the bright foreground star TYC~1967-1114-1 used for the AO correction is outside the scientific field of view, shifts between the individual frames were computed adopting as reference the centroid of the imaged 2MASS~J09563976+2900457 
foreground star (J=13.6, H=13.0) located $\sim$20$^{\prime\prime}$ west of DDO68~C (see panel e) of Fig.~\ref{fig_multi}).
Upon selection of the best images with FWHM$\leq$0.15$^{\prime\prime}$, individual exposures were combined into final stacked J and H deep images with total exposure times of $\sim$3864 s and $\sim$6944 s, respectively. 

 Fig.~\ref{fig_multi} presents a comparison between the LUCI AO final combined image in H and the seeing-limited SDSS $g$ image, the GALEX FUV image, and the  HI emission map of DDO68~C from \cite{Cannon2014}. While DDO68~C is clearly identified in the FUV and HI images, the galaxy is completely outshined by the bright TYC~1967-1114-1 star in the SDSS data; on the other hand, its individual stars are well resolved in the SOUL+LUCI AO images, with the vast majority of them located in the region of the strongest UV emission.     
 Portions of the J and H images zoomed on DDO68~C's  resolved stellar component are shown in Fig.~\ref{fig_lbt}.

\begin{figure*}
\begin{center}
\includegraphics[width=15cm]{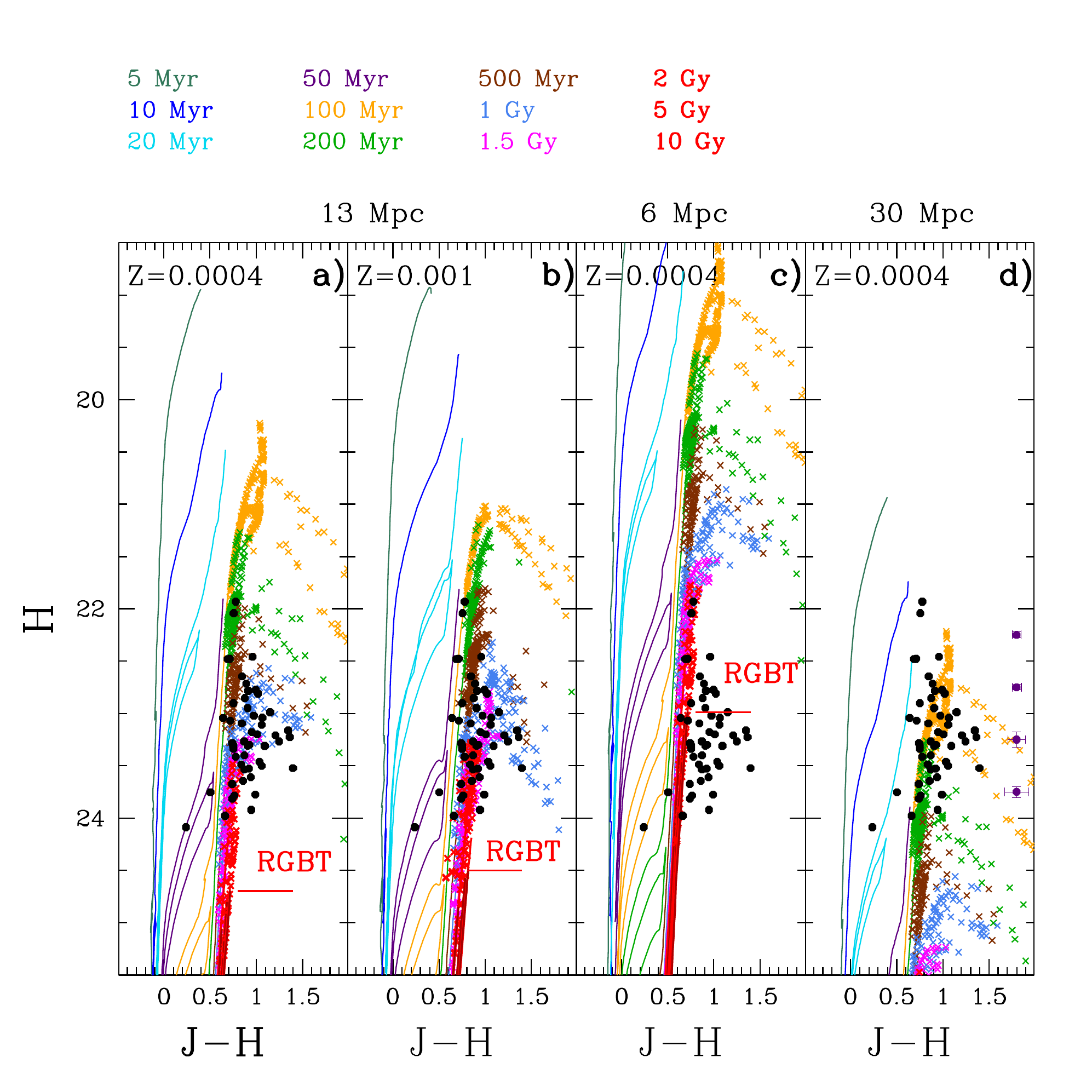}
    \caption{H vs. J-H CMD of stars (black dots) resolved in DDO68~C with LUCI AO. Superimposed as crosses with different colors, as indicated in the legend, are the PARSEC+COLIBRI stellar isochrones \citep{Bressan12,Marigo13,Rosenfield16} for different ages and for metallicities of   Z=0.0004 ($\sim$2.6\% solar) and  Z=0.001 ($\sim$6.6\% solar) shifted to  DDO68's distance of $\sim$13 Mpc (panels a and b). 
  At that distance, the photometry is compatible with stars $\sim$50 Myr to $\sim$2 Gyr old. Stars older than $\sim$2 Gyr 
 are too faint to be detected, with the expected magnitude of the RGB tip (RGBT) indicated by the horizontal segment. Panels c and d provide a comparison of
 the observed CMD with isochrones shifted to distances of 6 Mpc and 30 Mpc, respectively. Typical photometric errors in H and J-H as a function of magnitude are shown in the lower right portion of panel d.}
    \label{fig_cmd}
\end{center}
\end{figure*}

Aperture photometry with the PHOT task in the IRAF environment
\footnote{IRAF was distributed by the National Optical Astronomy Observatory, which was managed by the Association of Universities for 
Research in Astronomy (AURA) under a cooperative agreement with the National Science Foundation.}
was performed on the J and H stacked images using 
as input the position of sources detected above three times the background standard deviation in the deeper H image, and later  
allowing for a refined re-centering of the source positions through the ``centroid'' option. Source fluxes were measured within a 0.09$^{\prime\prime}$  diameter circular aperture with the local sky evaluated in an annulus at 0.3$^{\prime\prime}$ radius. 
Aperture corrections in J and H were evaluated from the brightest stars in DDO68~C, while the foreground star 
2MASS~J09563976+2900457 was used to derive photometric zeropoints (see Appendix~\ref{sec_appendix} for details). 
Eventually, the J and H catalogs were cross-correlated and combined into a final master catalog using the  CataXcorr and CataComb routines\footnote{Part of a package of astronomical softwares (CataPack) developed by P. Montegriffo at INAF-OABo.}. 
Spurious detections at the detector's edges and extended background galaxies were removed through visual inspection of the J and H images, providing a final catalog of $\sim$50 bona-fide stars with photometry in both bands.   
The photometric reduction of the LUCI AO data was repeated using SExtractor \citep{Bertin96}, including an independent estimate of the aperture corrections, in order to evaluate photometric uncertainties.  
The dispersion of the PHOT minus SExtractor magnitude and color difference distributions was used to infer average errors as a function of magnitude. 
The typical magnitude and color errors for a star as faint as  H$\sim$23.75 mag are $\sigma_H\sim0.05$ and $\sigma_{J-H}\sim0.13$. 
The derived uncertainties in different magnitude bins are shown in Fig.~\ref{fig_cmd}.

\section{Color Magnitude Diagrams} \label{cmd}

The H, J-H CMD of DDO68~C derived from the $\sim$50 stars resolved in the LUCI images is shown in Fig.~\ref{fig_cmd}. The CMD exhibits a sparse distribution of stars with colors in the range 0.2$\lesssim$J-H$\lesssim$1.4 and magnitudes of 
22$\lesssim$H$\lesssim$24. The CMD does not show the typical discontinuity associated with the red giant branch (RGB) tip detection 
used to infer galaxy distances \citep{Madore1995,Bellazzini2004}.  
Indeed, under the assumption that DDO68~C is located at the same $\sim$13 Mpc distance as DDO68, we would expect to detect the RGB tip  at H$\gtrsim$24.5 (the exact value depending on stellar population ages and metallicities), i.e., at least $\sim$0.5 mag fainter than our achieved  photometric depth. This is well shown in Fig.~\ref{fig_cmd}, where we compare the observed CMD with the PARSEC+COLIBRI stellar isochrones 
\citep{Bressan12,Marigo13,Rosenfield16} for different ages and for two metallicity values of Z=0.0004 and Z=0.001 
\citep[i.e. $\sim$2.6\% and $\sim$6.6\% solar for an adopted solar value of Z$_{\odot}$=0.0152 from][]{Caffau2011}.
For an assumed 13 Mpc distance, the observed CMD is in very good agreement with the distribution of bright asymptotic giant branch (AGB) stars 
predicted by isochrones $\sim$50 Myr to $\sim$2 Gyr old.   
Core helium-burning stars in the so-called blue loop phase, with an age of $\sim$50 Myr, are also present. Stars older than $\sim$2 Gyr are too faint to be detected. 

\begin{figure*}
\includegraphics[width=\textwidth]{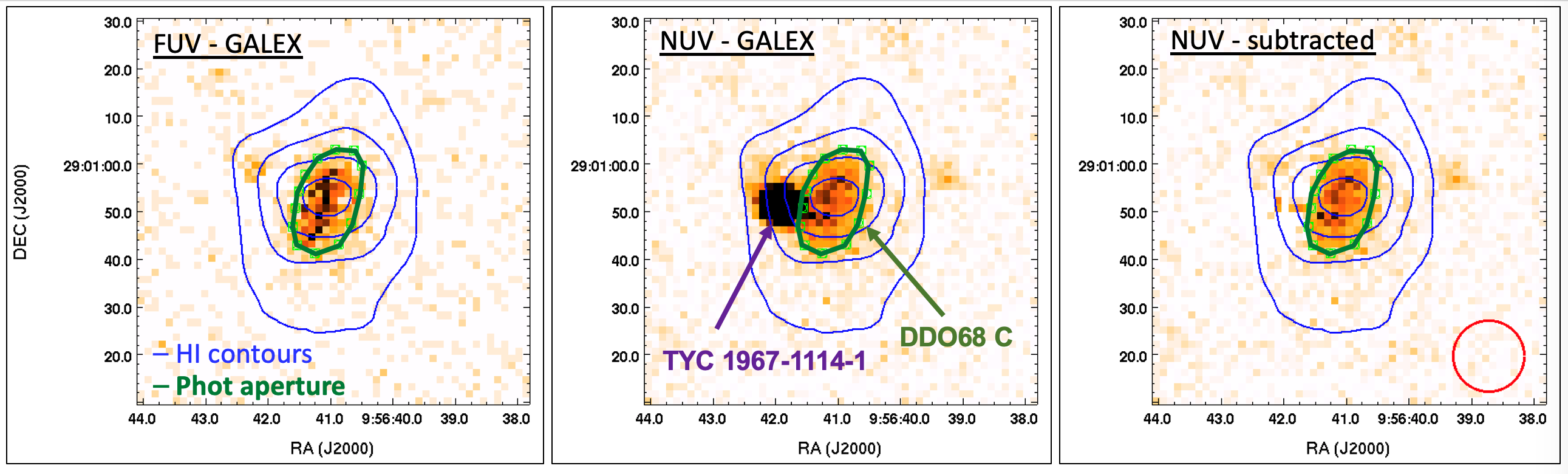}
    \caption{GALEX FUV and NUV images of DDO68~C with a field of view of 80$^{\prime\prime}\times$ 80$^{\prime\prime}$. Superimposed in blue, thin curves, are the HI contours at (2, 4, 6, 8) $\times$ 10$^{20}$\ cm$^{-2}$ from the same data of \cite{Cannon2014}. The thicker green polygon is the aperture used to compute the FUV and NUV magnitudes.  
    The bright foreground stars  TYC~1967-1114-1, well visible in the NUV central panel image, was subtracted in the rightmost panel before computing the galaxy NUV emission. Here the circle denotes the 15$^{\prime\prime}$ beam of the HI data.}
    \label{fig_galex}
\end{figure*}

For distances closer than 13 Mpc, the isochrones are shifted toward brighter magnitudes (see e.g. panel c for an assumed distance of $\sim$6 Mpc). 
In this scenario, the CMD is  poorly consistent with the stellar models, since the reddest stars with J-H$\gtrsim$1 are incompatible with the relatively blue colors of the RGB feature, considering also the relatively low photometric uncertainty of $\sigma_{J-H}\lesssim 0.1$ mag. 
The observed red colors could in principle  be reproduced by assuming metallicities of solar or higher, which is however unrealistic for the mass of DDO68~C (see section~\ref{discussion}) 
given the mass-metallicity relation of dwarf galaxies \citep{Berg12}. 

At the opposite extreme, the CMD allows us to infer upper limits on DDO68~C's distance. In panel d, the isochrones have been shifted to a distance of $\sim$30 Mpc to match the luminosity of the 10 Myr old models with the brightest measured star.  
This is the youngest population (and the farthest distance) allowed for DDO68~C based on the lack of significant H$\alpha$ emission \citep{Cannon2014}, excluding a major population of ionizing stars; star formation more recent than 10 Myr ago can also be excluded 
based on the integrated-light FUV-NUV color, as we discuss in the next section.

\section{Discussion and Conclusions} \label{discussion}

The new SOUL+LUCI AO data allowed us to resolve, for the first time ever, individual stars in the small gas rich galaxy DDO68~C. 
Although the photometry is not deep enough to derive a direct distance measurement through the RGB tip method, the CMD appears fully 
compatible with a scenario in which the system is at the same $\sim$13 Mpc distance as DDO68. Under this assumption, the comparison of the CMD with 
stellar evolution models reveals a population of stars with ages between $\sim$50 Myr and $\sim$2 Gyr, while the data are blind to older, fainter stars. Here we discuss how GALEX UV data can provide complementary information to constrain DDO68~C's stellar populations and distance.

\begin{figure*}
\begin{center}
\includegraphics[width=5.7cm]{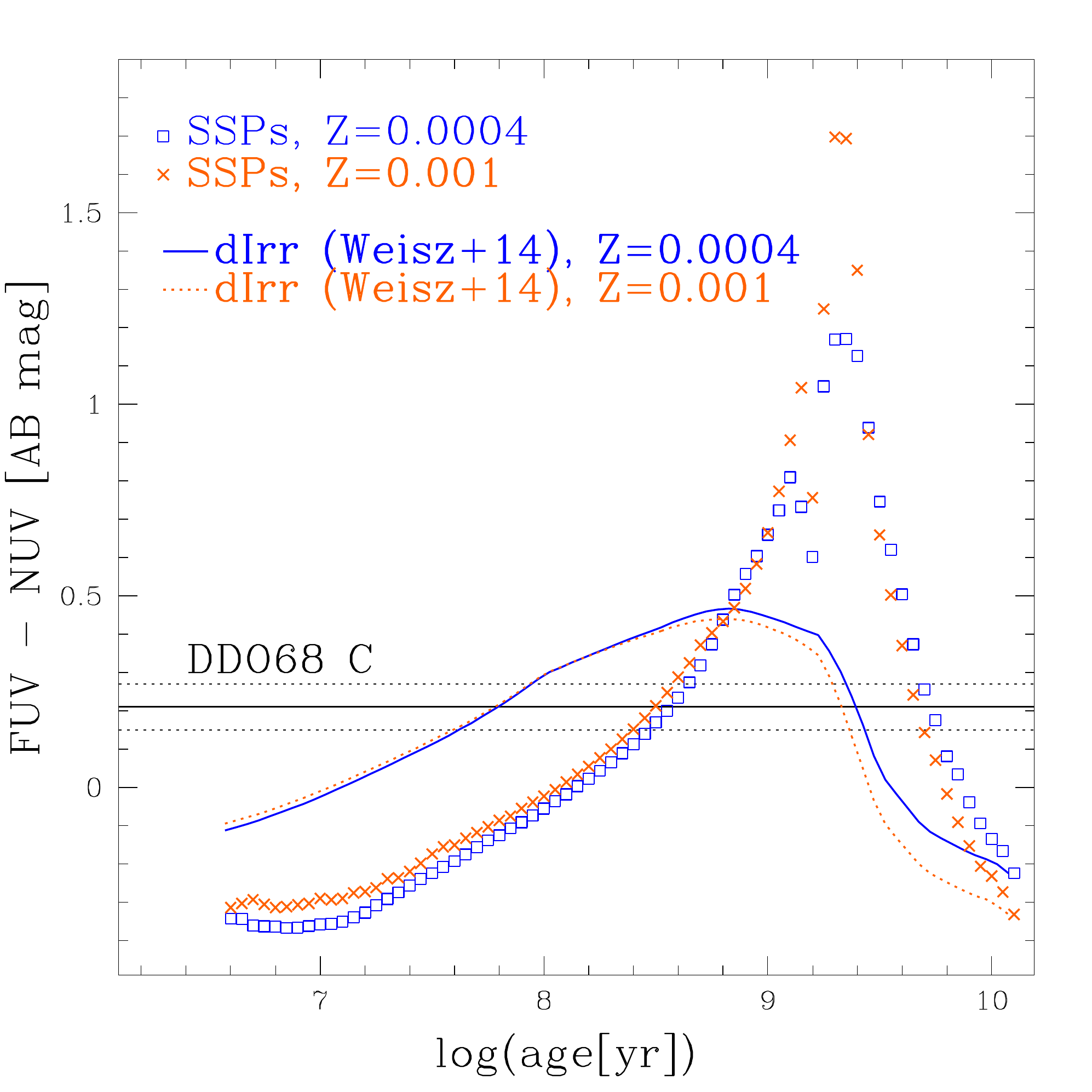}
\includegraphics[width=5.7cm]{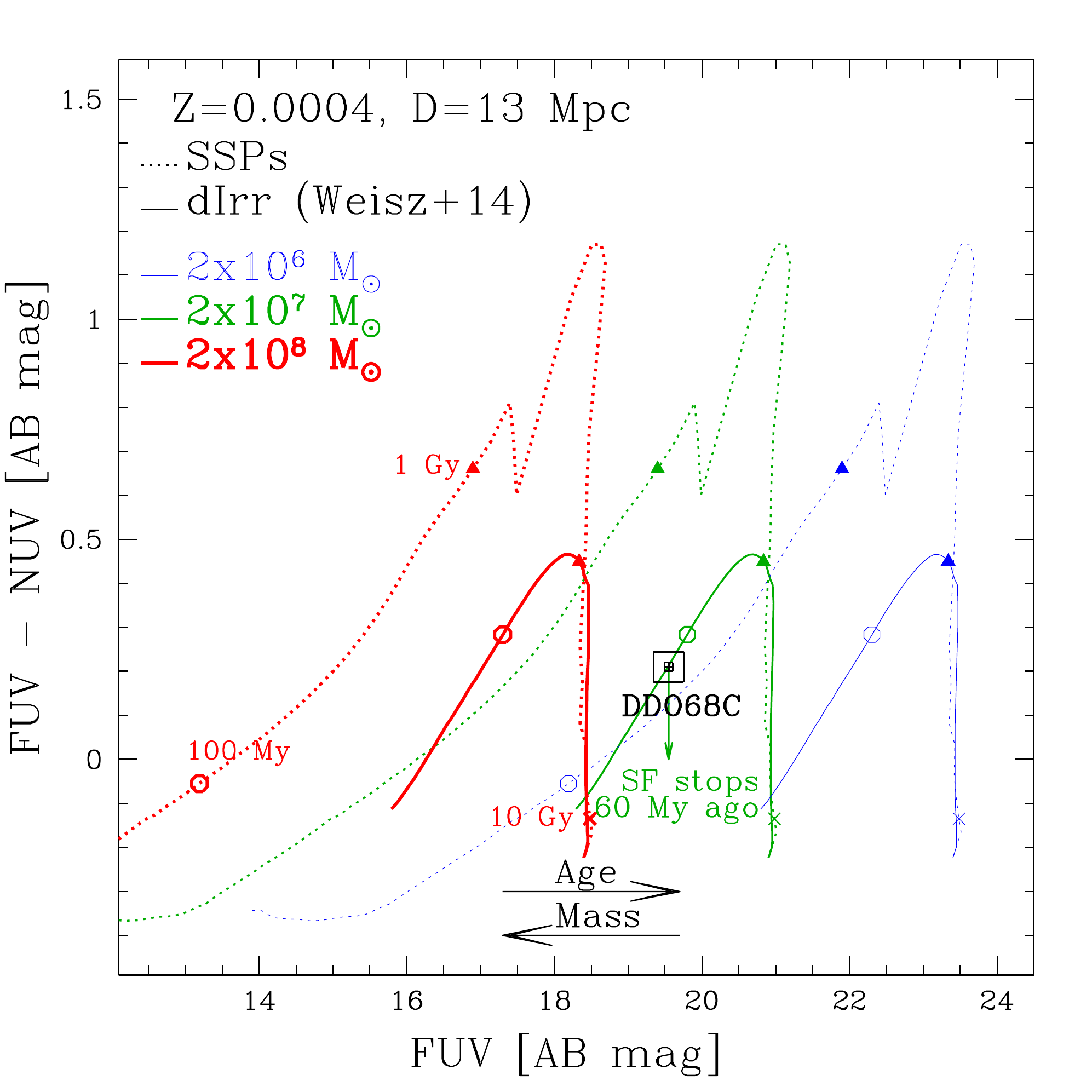}
\includegraphics[width=5.7cm]{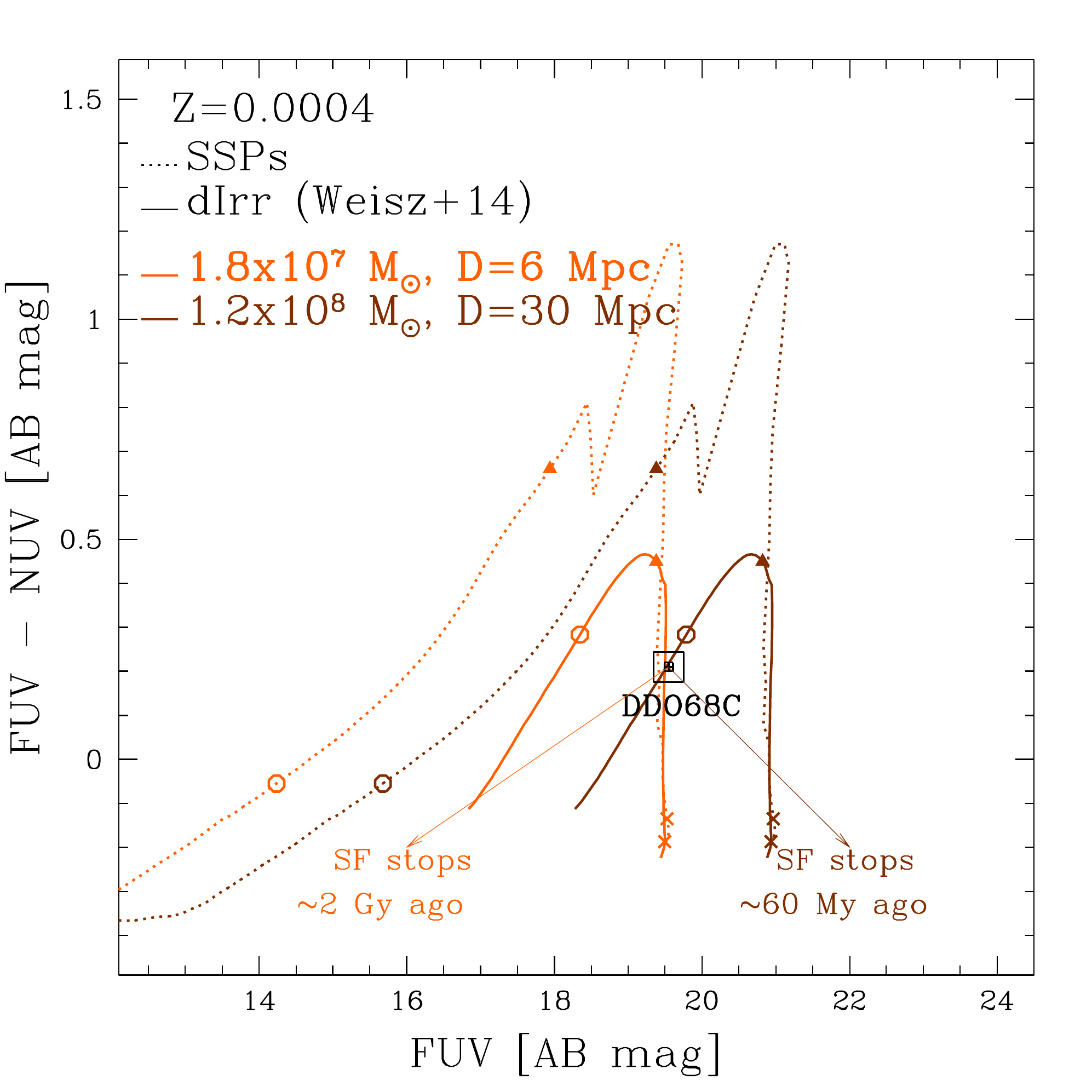}
    \caption{{\bf Left:} predicted FUV-NUV GALEX colors in the AB mag system as a function of stellar age for simple stellar populations (SSPs, open symbols) generated from the PARSEC+COLIBRI models for a \cite{Kroupa2001} IMF and for two metallicities of Z=0.0004 and Z=0.001. The solid and dashed curves correspond instead to a typical dIrr SFH \citep{Weisz14} with the assumption of different quenching epochs, indicated in the x axis (see text for details). The horizontal lines denote the average color and associated uncertainty for DDO68~C. {\bf Middle:} same models (only for Z=$0.0004$) in the FUV-NUV versus FUV plane, assuming a distance of 13 Mpc and different stellar masses of  2$\times$10$^6$, 2$\times$10$^7$, and 2$\times$ 10$^8$  M$_{\odot}$. Along each model, age increases from left to right. The observed values for DDO68~C (notice that the uncertainties are smaller than the symbol size) are compatible with a dIrr SFH assuming quenching at about $\sim$60 Myr ago and a total galaxy stellar mass of $\sim$2$\times$10$^7$  M$_{\odot}$. 
    {\bf Right:} two alternative solutions, with distances of 6 Mpc and 30 Mpc, stellar masses of $\sim$2$\times$10$^7$ and $\sim\times$ 10$^8$  M$_{\odot}$, and quenching epochs at $\sim$2 Gyr and $\sim$60 Myr ago, respectively, are also compatible with the GALEX magnitudes of DDO68~C.}
    \label{fig_ssp}
    \end{center}
\end{figure*}

We downloaded GALEX *int.fits images of DDO68~C, in units of counts sec$^{-1}$pixel$^{-1}$, from the MAST archive\footnote{https://mast.stsci.edu/portal/Mashup/Clients/Mast/Portal.html}. PSF-fitting photometry 
with the  DAOPHOT package \citep{Stetson87} in IRAF was run to subtract the bright foreground star TYC 1967-1114-1 in the NUV image (see Fig.~\ref{fig_galex}). 
Then aperture photometry was performed in both  FUV and NUV with the IRAF POLYPHOT task  
within a polygonal aperture of $\sim$250 arcsec$^2$  area encompassing the FUV contour at 
$\sim$0.6 $\times$10$^{-15}$ erg\ s$^{-1}$\ cm$^{-2}$ in Fig.~\ref{fig_multi}, enclosing the vast majority of the UV galaxy emission. 
Repeated flux measurements, obtained subtracting the background computed in different regions around DDO68~C, were 
combined to obtain average magnitudes and their standard deviations. To account for the uncertainty due to the removal of the foreground interloper in the NUV image, background values were 
estimated in the proximity of subtracted PSF stars. Final calibrated magnitudes in the AB system \citep{Oke1990} were derived through the relations 
FUV= -2.5$\times$log(counts/sec) + 18.82 and NUV= -2.5$\times$log(counts/sec) + 20.08 from the GALEX page  \footnote{https://asd.gsfc.nasa.gov/archive/galex/FAQ/counts\_background.html}. Foreground reddening corrections were applied adopting 
E(B-V)=0.016 from the NED\footnote{https://ned.ipac.caltech.edu/} and the extinction coefficients in the GALEX filters from \cite{Yuan2013}. 
In the end we obtain FUV$=19.55\pm0.01$ mag and FUV$-$NUV$=0.21\pm0.06$. 

Fig.~\ref{fig_ssp} provides a comparison between the observed DDO68~C ultraviolet emission and model predictions for both simple stellar populations (SSPs) and a complex star formation history (SFH) appropriate for dwarf irregular (dIrr) galaxies. 
The displayed SSPs are based on the PARSEC+COLIBRI stellar evolution models \citep{Bressan12,Marigo13,Rosenfield16} and were downloaded from the public web interface at http://stev.oapd.inaf.it/cgi-bin/cmd. 
Conversion from the models' Vegamag system to the AB system was performed through the relations FUV(AB)=FUV(Vega)+ 2.128 and NUV(AB)=NUV(Vega)+1.662 (L. Girardi, private communication).  
The displayed models are for a \cite{Kroupa2001} IMF, for the same metallicity values of Z=0.0004 and 0.001 as in Fig.~\ref{fig_cmd}, and span ages from $\sim$4 Myr to 13 Gyr. 
From young to progressively older ages, the FUV luminosity decreases monotonically. At the youngest ages, the SSPs have the bluest FUV-NUV colors
(of about -0.4 mag for a Z=0.0004 metallicity), because O-type and B-type stars highly contribute to the FUV flux; then the FUV-NUV color increases with age due to the decreasing temperature of turnoff stars. For ages older than $\sim$2 Gyr,  
the trend is inverted and the FUV-NUV color becomes progressively bluer 
due to the increased contribution from hot, post-AGB stars \citep[see][section~3, for a detailed discussion of the stars contributing to the FUV-NUV trend]{Rampazzo2011}.

To account for the more complex SFH of dIrrs, we combined the PARSEC+COLIBRI SSPs according to the behaviour of the median SFH derived by \cite{Weisz14} from 
a sample of Local Group dIrrs with deep HST CMDs. This median SFH starts as early as $\sim$13 Gyr ago and then proceeds almost continuously until present, with a typical rate which is about two times larger over the past $\sim$4 Gyr compared to earlier epochs. Starting from the \cite{Weisz14} SFH, we assumed quenching time to be a free-parameter; in Fig.~\ref{fig_ssp}, left panel, the values along the x axis correspond to the epoch when star formation stopped. For a still active star formation, the  FUV-NUV color is as blue as $\sim-0.1$, while a quenching occurred $\sim$1 Gyr ago produces a red color of $\sim$0.5.  The trend is then inverted for 
earlier termination epochs.

The left panel of Fig.~\ref{fig_ssp} shows that DDO68~C's color of FUV-NUV=0.21$\pm$0.06 is compatible, within the errors and for the adopted metallicities, with an SSP age between 250 Myr and 400 Myr. However, the CMD in Fig.~\ref{fig_cmd} suggests a spread in age. Under the assumption of the more complex dIrr SFH case, the UV properties of 
DDO68~C can be reproduced assuming quenching at $\sim$60 Myr ago, a scenario also in very good agreement with the CMD of  Fig.~\ref{fig_cmd} where the youngest populated isochrone is the 50 Myr old one. This solution provides a total stellar mass of $\sim$2$\times10^7$ \ M$_{\odot}$ for DDO68~C, as shown in the middle panel of Fig.~\ref{fig_ssp}.
Notice that the relatively red FUV-NUV color of DDO68~C  excludes the presence of 
current star formation, consistent both with the absence of stars younger than $\sim$10 Myr in the CMD, and with the lack of significant  H$\alpha$ emission \citep{Cannon2014}.

 We notice in  Fig.~\ref{fig_ssp} that the FUV-NUV color decrease for models older (or quenching epochs earlier) than 1-2 Gyr provides an alternative solution where all stars in DDO68~C are 
older than 2 Gyr.  
However, this scenario would require DDO68~C to lie at a much closer distance of $\sim$6 Mpc in order for the brightest resolved stars in the CMD to be compatible with the maximum luminosity of the $\sim$ 2 Gyr old stellar isochrones, as illustrated in panel c of Fig~\ref{fig_cmd}. We show in the right panel of Fig.~\ref{fig_ssp} that this scenario is compatible with the observed FUV and NUV galaxy magnitudes if a total stellar mass of $\sim$2$\times10^7$  \ M$_{\odot}$ is assumed. 
However, at a 6 Mpc distance, this mass should have formed within a 0.21 kpc$^2$ region, implying an average stellar mass density of $\Sigma_{\star}\sim$100 M$_{\odot}$pc$^{-2}$, which appears unrealistically high compared to values derived in other dIrrs
\citep[e.g.][]{Sacchi16,Mcquinn21}. 

At the opposite extreme, we explored a scenario in which DDO68~C lies at a much larger distance of $\sim$30 Mpc, the maximum distance allowed by the brightest resolved star in the CMD of Fig.~\ref{fig_cmd} given the lack of significant H$\alpha$ emission,
thus of ionizing stars younger than $\sim$10 Myr. According to the models in the right panel of Fig.~\ref{fig_ssp}, the galaxy stellar mass would be, in this case, as high as $\sim$10$^8$ M$_{\odot}$, potentially close to the galaxy dynamical mass. 
In addition, the relatively red FUV-NUV color of  DDO68~C appears incompatible with the presence of stars as young as 10 Myr, which would produce typically bluer colors.


In conclusion, the combined analysis of the SOUL+LUCI AO resolved-star photometry, archival GALEX data, and H$\alpha$ images from \cite{Cannon2014} support a scenario in which DDO68~C has been actively forming stars in the past (possibly since $\sim$13 Gyr ago, as in Local Group dIrrs) but has quenched (or significantly suppressed) its activity about 50-60 Myr ago.
All data, including the systemic HI velocity derived by \cite{Cannon2014}, are best reproduced by the assumption that  DDO68~C is at the same $\sim$13 Mpc distance as its candidate interacting companion DDO68.  
In fact, although the new AO data do not provide a direct distance measurement to DDO68~C through the RGB tip method
  (which would require the superb depth and spatial resolution of HST or JWST), the assumption of 
alternative distances tend to produce either a mismatch between the resolved-star CMD and stellar isochrones, and/or unrealistically high stellar mass or stellar mass density, and/or major inconsistency between the UV properties of DDO68~C and predictions from population synthesis models.  

 The proposed scenario in which  DDO68~C, DDO68~B, and S1 are all satellites of DDO68 puts new constraints on 
theoretical predictions on the number of satellites around isolated dwarf galaxies with the mass of DDO68. 
For instance, some cosmological N-body simulations coupled with abundance matching models foresee at most one luminous satellite more massive than 
M$_{\star} \sim 10^6$ M$_{\odot}$ around a host with the mass of DDO68 \citep{Dooley2017,Santos2022}.  
DDO68 is indeed the only case so far of a low mass dwarf (less massive than the Magellanic Clouds) 
known to be interacting with three smaller systems, a peculiarity that offers tantalizing explanation 
for its  anomalous extremely low metallicity.  
Although the actual gas mass of DDO68~C equals just few percents of the gas mass in DDO68, 
thus insufficient to dilute the gas to the extremely low metallicity observed \citep{Pascale2022}, 
it may have been higher in the past before being deposited into its more massive companion.
Deeper data on the diffuse gas connecting the two systems as well as more accurate 
dynamical mass estimates for DDO68~C would be crucial to better constrain hydrodynamical N-body 
simulations for the merging history of DDO68 with its satellites' population. 

\newpage

\acknowledgments

We thank L. Girardi for his precious support with the PARSEC+COLIBRI stellar models. 
 We are grateful to the anonymous referee for his/her constructive report that helped to improve the paper.
We acknowledge the support from the LBT-Italian Coordination Facility for the execution of observations, data distribution, and reduction. 
FA, MB, RP and MT acknowledge the financial support from INAF Main Stream grant 1.05.01.86.28 ``SSH''. 
F. A., L.K.H, and M.T. acknowledge funding from INAF PRIN-SKA-2017 program 1.05.01.88.04.

Based on data acquired using the Large Binocular Telescope (LBT). The LBT is an international collaboration amongst institutions in the United States, Italy, and Germany. LBT Corporation partners are The University of Arizona on behalf of the Arizona university system; Istituto Nazionale di Astrofisica, Italy; LBT Beteiligungsgesellschaft, Germany, representing the Max-Planck Society, the Astrophysical Institute Potsdam, and Heidelberg University; The Ohio State University; and The Research Corporation, on behalf of The University of Notre Dame, University of Minnesota, and University of Virginia.

This research has made use of the SIMBAD database, operated at CDS, Strasbourg, France.
This research has made use of the NASA/IPAC Extragalactic Database (NED) which is operated by the Jet Propulsion Laboratory, California Institute of Technology, under contract with the National Aeronautics and Space Administration. 
This research has made use of NASA's Astrophysics Data System.

%

\vspace{5mm}
\facilities{LBT(SOUL+LUCI), GALEX}


\software{ 
          SExtractor \citep{Bertin96},
            DAOPHOT \citep{Stetson87}  
            }

\appendix

\section{Spatially variable aperture corrections and photometric zeropoints} \label{sec_appendix}

Phootmetric calibration in the J and H bands for stars in DDO68~C was performed according to the formula:

\begin{equation}
{\rm J,H_{cal} = J,H_{6} + AC_{J,H} + ZP_{J,H},}
\label{eq_a1}
\end{equation}

where J,H${\rm _{cal}}$ are the calibrated magnitudes, J,H$_{6}$ are the instrumental magnitudes derived from aperture photometry within 
the adopted 6 pixel  (0.09$^{\prime\prime}$) diameter aperture, ${\rm AC_{J,H}}$ are the aperture corrections to a larger 5$\times$FWHM aperture (enclosing $\sim$80\% of the total flux), and  ${\rm ZP_{J,H}}$ are the photometric zeropoints in the two bands.  

The variation of the shape of the PSF with increasing distance from the AO star \citep[e.g.][]{Wilson1996} requires in fact the implementation of a spatially variable aperture correction. Indeed, for the bright foreground star 2MASS~J09563976+2900457, at $\sim$30$^{\prime\prime}$ distance from the AO star, we measure a PSF FWHM$\sim$0.15$^{\prime\prime}$ in J and $\sim$0.12$^{\prime\prime}$ in H;
on the other hand, stars belonging to DDO68~C at $\sim$10$^{\prime\prime}$ distance from the AO star have 
FWHM$\sim$0.07$^{\prime\prime}$ in J and $\sim$0.06$^{\prime\prime}$ in H. Notice that the larger PSF derived in J reflects the typically worse atmospheric conditions during observations in this band. 

\begin{figure}
\begin{center}
\includegraphics[width=8cm]{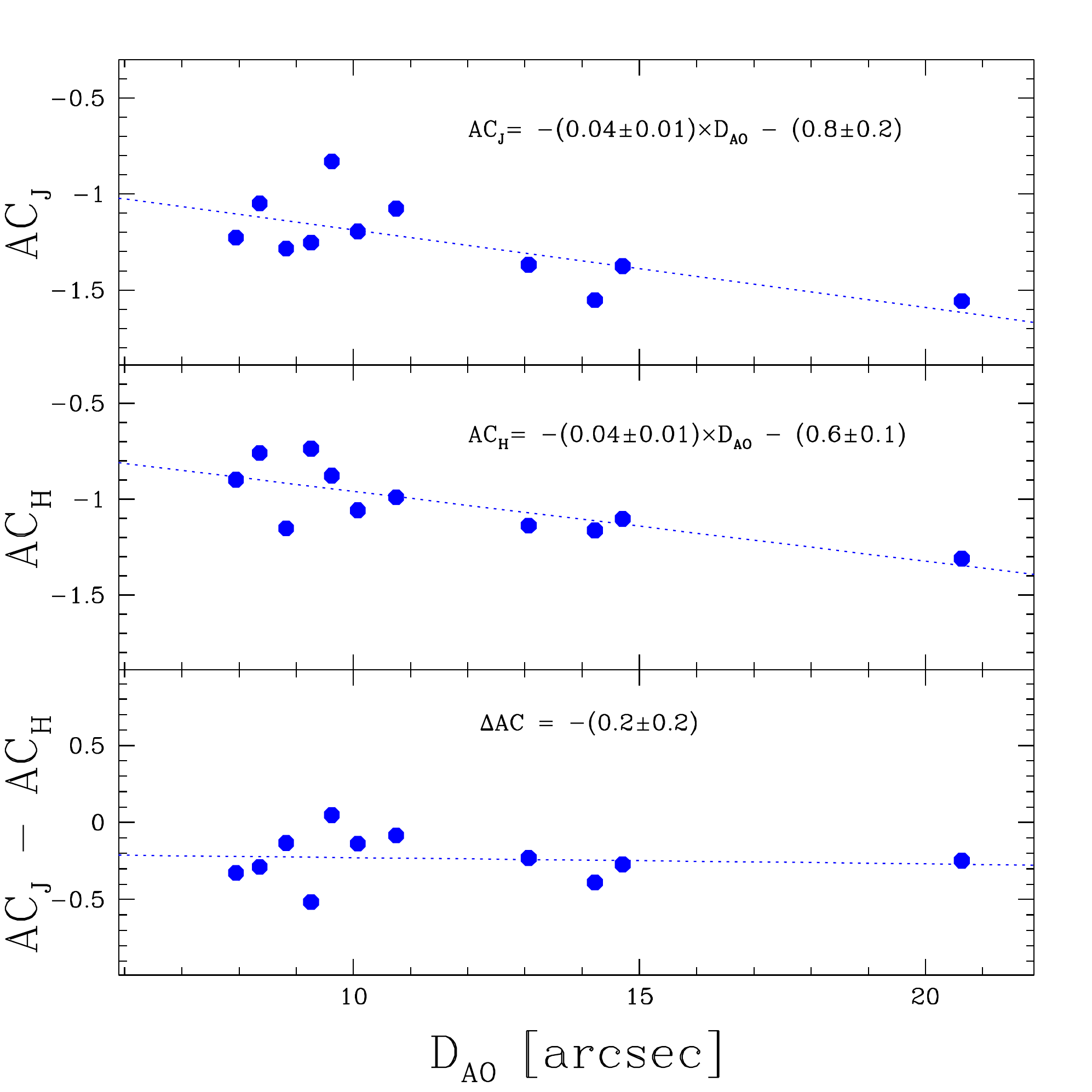}
\end{center}
\caption{Aperture corrections in J (${\rm AC_J}$) and H (${\rm AC_H}$) as a function of distance from the AO star 
derived from the brightest and most isolated stars in DDO68~C. The corrections were computed as the magnitude difference at apertures of  20/23 pixels (i.e., 5 times the FWHM at $\sim$10$^{\prime\prime}$) and 6 pixels. The bottom panel displays the difference between the J and H aperture corrections, showing no significant trend with distance from the AO star. Dotted lines are the least squares linear fits. 
} \label{fig_app}
\end{figure}

Aperture corrections were computed selecting the brightest (H$\lesssim$23) and most isolated stars in the observed field, paying attention to remove sources on top of prominent AO star spikes. We ended up with a sample of 11 stars, for which photometry was computed within a larger diameter aperture  of 
5 times the FWHM (i.e., 0.35$^{\prime\prime}$ or 23 pixels in J and  0.3$^{\prime\prime}$ or 20 pixels in H),  chosen to capture the largest possible fraction of the total flux while minimizing  contamination from background noise. Fig.~\ref{fig_app} shows the behaviour of the J and H aperture corrections (computed as the difference between the 20 or 23 pixel aperture photometry and the 6 pixel aperture photometry) as a function of distance from the AO star. 
As expected, these corrections increase (in absolute value) with distance from the AO star due to the increase in the PSF FWHM. However, because of the 
similar slopes of the relations in J and H, the trend on the combined J-H color is flat, as shown in the bottom panel of  Fig,~\ref{fig_app}. 

Least squares linear fits to the data points provide the following relations for the spatially variable aperture corrections: 

\begin{equation}
{\rm AC_J= mag_{J,5FWHM} - mag_{J,6} = -0.04 \times D_{AO}- 0.8} \label{a2}
\end{equation}

\begin{equation}
{\rm AC_H= mag_{H,5FWHM}-mag_{H,6} = -0.04 \times D_{AO}- 0.6} \label{a3}
\end{equation}

where D$_{AO}$ is the distance from the AO star in arcsec. For each star in DDO68~C, aperture corrections were then calculated injecting its AO star distance into Eqs.~\ref{a2} and ~\ref{a3}.

Photometric zeropoints were derived using the bright foreground star 2MASS~J09563976+2900457, with total magnitudes of  
J=13.60$\pm$0.02 and H=13.00$\pm0.03$ from \cite{2Mass}. Photometry in J and H was then performed for this star from our images within diameter  apertures of 0.75$^{\prime\prime}$ and 0.6$^{\prime\prime}$ in J and H,  respectively (i.e., 5 times its FWHM), in order to properly tie the zeropoints to the aperture corrections provided by Eq.~\ref{a2} and ~\ref{a3}. 

The derived zerpoints are:

\begin{equation}
{\rm ZP_J= J_{2MASS} - J_{5 FWHM} = 28.32}
\end{equation}

\begin{equation}
{\rm ZP_H= H_{2MASS} - H_{5 FWHM} = 28.03.}
\end{equation}

The  uncertainties in the overall photometric calibration are of order 0.1-0.2 mag (see the uncertainties associated to the zeropoints 
of the relations in Fig.~\ref{fig_app}) and they do not change significantly our conclusions.



\end{document}